\documentclass[aps,prl,reprint,superscriptaddress]{revtex4-2}
\usepackage[utf8]{inputenc}
\usepackage{graphicx}% Include figure files
\usepackage{dcolumn}% Align table columns on decimal point
\usepackage{bm}% bold math
\usepackage{amsmath}

\usepackage[separate-uncertainty = true, multi-part-units=single]{siunitx}
\usepackage[utf8]{inputenc}
\usepackage[T1]{fontenc}
\usepackage{mathptmx}
\usepackage{xcolor}
\usepackage{lineno}

\usepackage[normalem]{ulem}
\usepackage{verbatim}

\begin{document}

%\preprint{}

\title{Telecom wavelength quantum dots interfaced with silicon-nitride circuits via photonic wire bonding} %Title of paper



\author{Ulrich Pfister}\thanks{Contributed equally to this work}
\email[]{ulrich.pfister@ihfg.uni-stuttgart.de}
\affiliation{Institut für Halbleiteroptik und Funktionelle Grenzflächen, Center for Integrated Quantum Science and Technology (IQ$^{ST}$) and SCoPE, University of Stuttgart, Allmandring 3, 70569 Stuttgart, Germany}

\author{Daniel Wendland}
\thanks{Contributed equally to this work}
\email[]{daniel.wendland@uni-muenster.de}
\affiliation{Institute of Physics and Center for Nanotechnology, University of M\"unster, 48149 M\"unster, Germany}
\affiliation{Kirchhoff-Institute for Physics,
University of Heidelberg, 69120 Heidelberg, Germany}

\author{Florian Hornung}
\affiliation{Institut für Halbleiteroptik und Funktionelle Grenzflächen, Center for Integrated Quantum Science and Technology (IQ$^{ST}$) and SCoPE, University of Stuttgart, Allmandring 3, 70569 Stuttgart, Germany}

\author{Lena Engel}
\affiliation{Institut für Halbleiteroptik und Funktionelle Grenzflächen, Center for Integrated Quantum Science and Technology (IQ$^{ST}$) and SCoPE, University of Stuttgart, Allmandring 3, 70569 Stuttgart, Germany}
\author{Hendrik H\"uging}
\affiliation{Institute of Physics and Center for Nanotechnology, University of M\"unster, 48149 M\"unster, Germany}
\author{Elias Herzog}
\author{Ponraj Vijayan}
\author{Raphael Joos}
\affiliation{Institut für Halbleiteroptik und Funktionelle Grenzflächen, Center for Integrated Quantum Science and Technology (IQ$^{ST}$) and SCoPE, University of Stuttgart, Allmandring 3, 70569 Stuttgart, Germany}
\author{Erik Jung}
\affiliation{Kirchhoff-Institute for Physics, University of Heidelberg, 69120 Heidelberg, Germany}
%{Department of Chemistry, Second University, Nearby Town}

\author{Michael Jetter}

\author{Simone L. Portalupi}
\affiliation{Institut für Halbleiteroptik und Funktionelle Grenzflächen, Center for Integrated Quantum Science and Technology (IQ$^{ST}$) and SCoPE, University of Stuttgart, Allmandring 3, 70569 Stuttgart, Germany}

\author{Wolfram H. P. Pernice}
\affiliation{Institute of Physics and Center for Nanotechnology, University of M\"unster, 48149 M\"unster, Germany}
\affiliation{Kirchhoff-Institute for Physics,
University of Heidelberg, 69120 Heidelberg, Germany}

\author{Peter Michler}
\affiliation{Institut für Halbleiteroptik und Funktionelle Grenzflächen, Center for Integrated Quantum Science and Technology (IQ$^{ST}$) and SCoPE, University of Stuttgart, Allmandring 3, 70569 Stuttgart, Germany}

\begin{abstract}
Photonic integrated circuits find ubiquitous use in various technologies, from communication, to computing and sensing, and therefore play a crucial role in the quantum technology counterparts. 
Several systems are currently under investigation, each showing distinct advantages and drawbacks. 
For this reason, efforts are made to effectively combine different platforms in order to benefit from their respective strengths. In this work, 3D laser written photonic wire bonds are employed to interface triggered sources of quantum light, based on semiconductor quantum dots embedded into etched microlenses, with low-loss silicon-nitride photonics. Single photons at telecom wavelengths are generated by the In(Ga)As quantum dots which are then funneled into a silicon-nitride chip containing single-mode waveguides and beamsplitters. 
The second-order correlation function of $g^{(2)}(0) = \SI{0.11\pm0.02}{}$, measured via the on-chip beamsplitter, clearly demonstrates the transfer of single photons into the silicon-nitride platform. 
%An average wire bonding efficiency of \SI{11.19\pm3.97}{\percent} can be achieved even at cryogenic temperatures, 
The photonic wire bonds funnel on average \SI{28.6(8.8)}{\percent} of the bare microlens emission ($\text{NA}=0.6$) into the silicon-nitride-based photonic integrated circuit even at cryogenic temperatures. This opens the route for the effective future up-scaling of circuitry complexity based on the use of multiple different platforms.
\end{abstract}
\pacs{}% insert suggested PACS numbers in braces on next line

\maketitle %\maketitle must follow title, authors, abstract and \pacs

\section{Introduction}
For modern quantum technologies, the transfer from large bulk optics experiments \cite{Brien2003,Zhong2020} to scalable
photonic chips with several optical elements is highly desirable \cite{Wang2020}. Indeed, several implementations in quantum communication \cite{Lo2014}, optical quantum computing \cite{Maring2024}, simulation \cite{Somhorst2023}, and sensing \cite{Stokowski2023} can strongly benefit from the possibility to increase the experimental complexity still on a small footprint device. 
In 2001, an efficient photonic quantum computation scheme was proposed by using only linear optics elements, single-photon sources, beamsplitters, phase shifters, and photodetectors \cite{Knill2001}. This stimulated the search for a platform capable of including all these elements simultaneously, ideally providing state-of-the-art performance for each component. Silicon (Si) and silicon-nitride (Si$_3$N$_4$) showed that large-scale photonic integrated circuits can be realized, thanks to the achievable low-loss in light propagation\cite{Wang2020}. For the generation of quantum light, spontaneous four-wave mixing or spontaneous parametric down conversion have been used in exciting experiments \cite{Spring2017,Kaneda2019,Wang2018}. Nevertheless, the probabilistic nature of the light emission process can impact the further upscaling of the experimental complexity. This is where deterministic sources of quantum light, as for example semiconductor quantum dots (QDs), can play a key role \cite{MichlerPortalupi+2024}. In 2018, the simultaneous operation of a III-V chip with QDs as single-photon source, single-mode waveguides (WGs), a 50:50 beamsplitter and two-single-photon detectors was demonstrated \cite{Schwartz2018}. Still, the high losses observed in the photonic circuitry represent a challenge in reaching the same photonic complexity achieved in Si and related material systems. Therefore, it seems advantageous to combine these platforms in order to benefit from each other’s key strengths using a hybrid interface \cite{Elshaari2020}. 
There already exist several approaches including wafer bonding \cite{Davanco2017,Schnauber2019,Vijayan2024}, transfer printing \cite{Osada2019,Katsumi2023} and pick and place techniques \cite{Elshaari2017,Elshaari2018}, combining the advantages of different platforms. 
\begin{figure}[hbt!]
\includegraphics[width=0.45\textwidth]{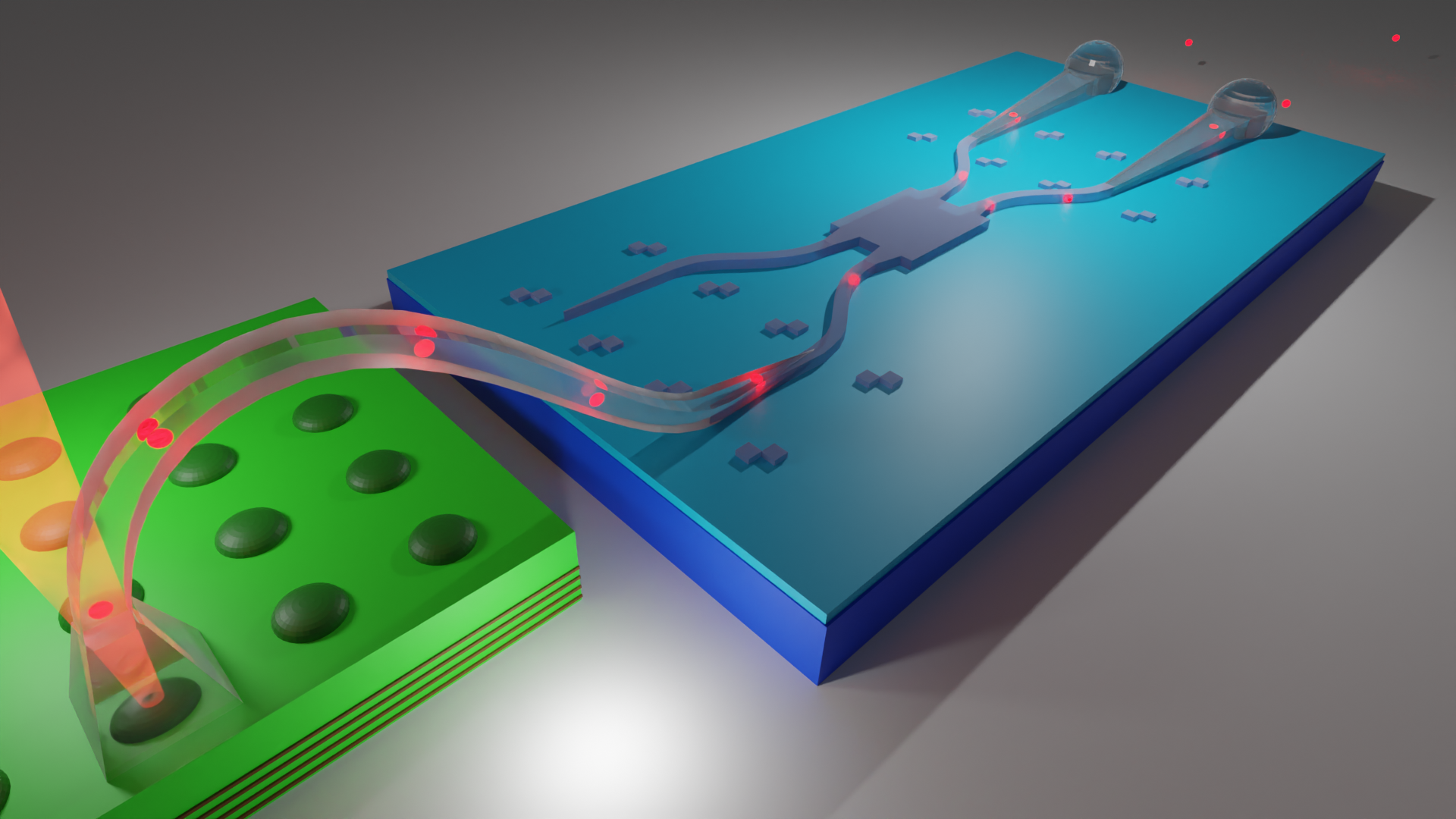}
\caption{\label{fig:Excitation_Sketch} Sketch of the hybrid photonic chip. On the left (in green), the InAs/GaAs platform is shown with the AlAs/GaAs DBRs and the microlenses above the quantum dot layer. On the right (in blue), the Si$_3$N$_4$-based platform with the corresponding fabricated beamsplitter is depicted. Both platforms are connected by the laser written photonic wire bonds. The top excitation of the QDs is also visualized.}
\end{figure}
For well established photonic integrated circuits (PICs) on low-loss platforms, 3D direct laser writing (DLW) \cite{Serbin2003,Guo2001,Sartison2021,Gehring2020,Johann2023} offers a high design flexibility to fabricated photonic elements on a micrometer scale providing versatile solutions for individual challenges. This flexibility has been already demonstrated via laser-written photonic wire bonds (PWB) \cite{Schumann2014} that allowed to guide laser emission into a PIC \cite{Billah2018}, a connection of two silicon-on-insulator PICs \cite{Lindenmann2012} and recently the guiding of single-photons from QDs, emitting in the near-infrared embedded in monolithically etched WGs, into a fiber array \cite{Gregorio2024}. 
In this work PWBs are employed to form an interface between the III-V platform and the Si$_3$N$_4$-based PIC for an on-chip beamsplitting, as schematically shown in Figure \ref{fig:Excitation_Sketch}. Indium gallium arsenide (In(Ga)As) QDs are used as single-photon sources emitting at telecom-wavelengths in combination with truncated Gaussian-shaped microlenses for a more Gaussian-like emission profile.
\begin{figure*}[hbt]
\includegraphics{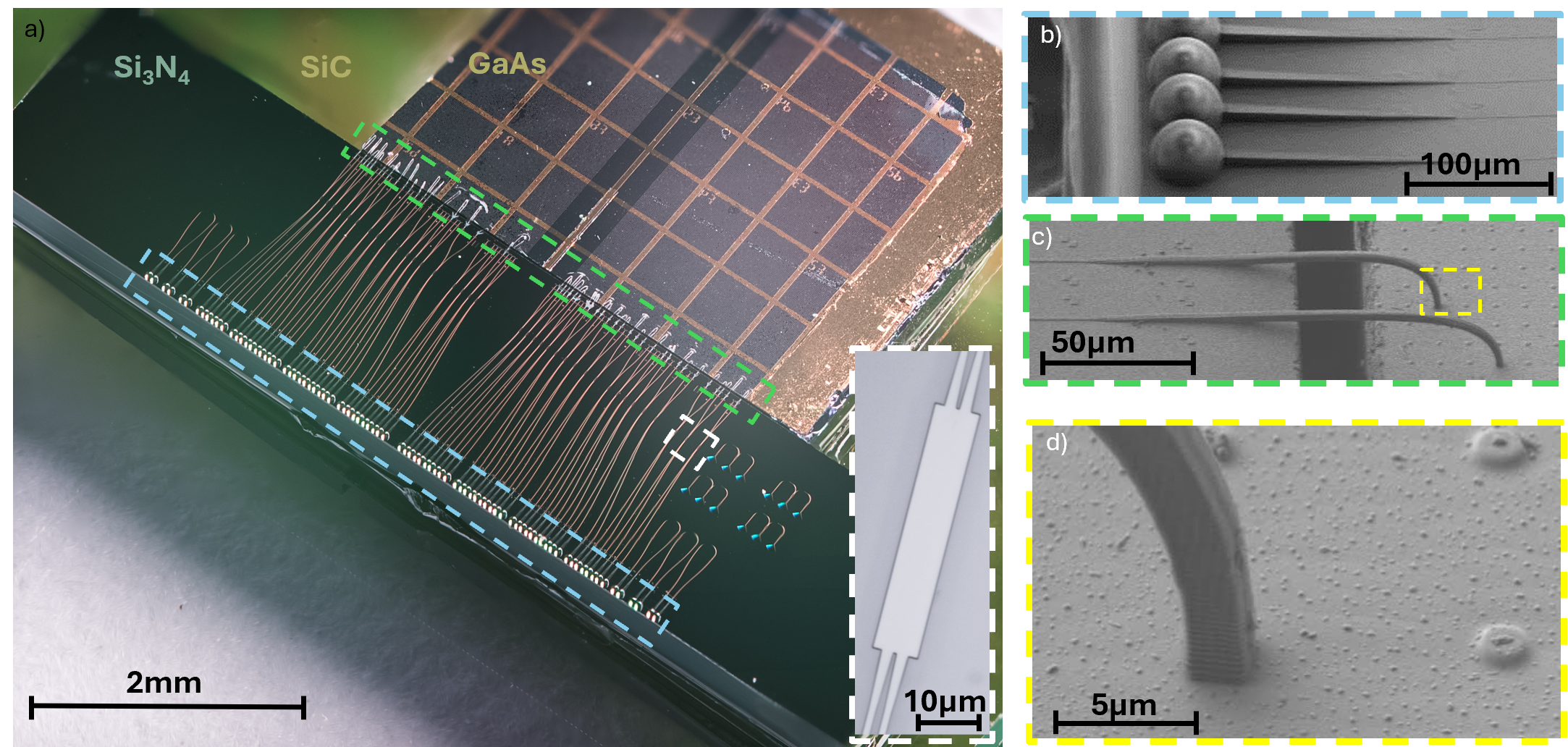}
\caption{\label{fig:chip_figure}a) The photograph shows the fully assembled hybrid system, consisting of the In(Ga)As/GaAs QD sample, which is aligned to a PIC sample made from Si$_3$N$_4$. The aligned samples are glued to a SiC carrier chip. In the inset, a microscope picture of the fabricated MMIs can be seen. b) The scanning electron micrograph shows the side couplers that couple the light from the chip to the collection objective. They are fabricated using DLW, and consist of a linear taper and a spherical lens. c) Photonic wire bonds bridge the gap of roughly \SI{30}{\micro\meter} between both samples. On the left the PIC sample can be seen with WGs and alignment markers made from Si$_3$N$_4$. d) The highlighted area from c) is shown as a close up. On the right two microlenses can be seen that have not been contacted, while on the left the interface of a PWB and a microlens is shown.
}
\end{figure*}

\section{Device Fabrication}
The sample containing the QDs was grown on a GaAs substrate via metal-organic vapor-phase epitaxy. It consists of a distributed Bragg reflector (DBR) formed by 23 AlAs/GaAs pairs acting as a bottom mirror to reduce losses into the substrate. Subsequently, an InGaAs metamorphic buffer (MMB) \cite{Sittig2022} was used, acting as a strain relaxation layer for engineering of the emission wavelength to the C-/S- band. Next, the In(Ga)As QD layer followed and was capped afterwards by a \SI{805}{\nano\meter} thick InGaAs layer which is needed for the successive fabrication of the microlenses. These truncated Gaussian-shaped microlenses, designed to enhance the photon extraction efficiency, were fabricated using an established wet-chemical etching process \cite{Sartison2018}. 
Using FDTD, the target dimensions of the micorlens with a diameter of $\SI{2.2}{\micro \meter}$, lens height of \SI{350}{\nano\meter} and a distance of \SI{400}{\nano\meter} from the QD layer to the microlens baseline, were determined for optimum performance.
After the etching, atomic force microscopy measurements were performed to identify a region close to these ideal parameters. Once this region was located, the sample was cleaved to ensure that the microlenses were found close to the chip edge. 
In this way, the distance between the III-V sample and the Si$_3$N$_4$-based PIC is minimized. This reduces the necessary PWB printing length, improving the mechanical stability as well as the propagation losses.
Before the actual printing, the array of lenses were pre-characterized with a free-space micro-photoluminescence setup employing an objective with a numerical aperture (NA) of 0.6 to find suitable candidates for the PWB. This allows to compare the behavior of the non-deterministically positioned microlenses before and after the 3D printing, being particularly important for the reliable quantification of the efficiency of the approach.
The initial characterization of the etched microlenses was performed in a helium-flow cryostat at 4 K with a confocal setup configuration. All mircolenses in the relevant region were investigated and the corresponding spectra of the QDs were measured under above-band (AB) excitation in saturation with a continuous wave laser at \SI{800}{\nano\meter}. In total, 52 suitable candidates
%an average emission wavelength of $\SI{1503.81(2.05)}{\nano\meter}$
were found and 47 of them showed sufficient separation.
Next, the Si$_3$N$_4$-based PIC was designed according to the positions of the pre-selected microlenses. This was done with the help of the open-source python-based framework gdshelpers \cite{Gehring2019}. The integrated structures were designed so that their positions match those of the pre-characterized lenses. 
For the fabrication of the PIC, a \SI{525}{\micro \meter} thick Si handle wafer with a \SI{3.3}{\micro \meter} SiO$_2$ buffer layer and \SI{330}{\nano \meter} of stoichiometric low pressure chemical vapor deposited (LPCVD) Si$_3$N$_4$ was used. Since the III-V sample embedding the QDs had an overall thickness of \SI{380}{\micro \meter}, the Si$_3$N$_4$ chip was polished to reduce the bottom Si substrate to an overall thickness close to the III-V sample. This was done to reduce the necessary upwards and downwards bends in the bridging PWBs. 
In a next step, the PIC design was transferred on to the Si$_3$N$_4$ layer, using a negative tone resist, spin coating and electron beam lithography. Consecutively executed reactive ion etching formed the PIC in the Si$_3$N$_4$ layer. The PIC consists of multi-mode interference (MMI) couplers with two input and two output WGs (2x2) as shown in the inset in Figure \ref{fig:chip_figure}a). The input WGs are routed to the edge facing to the QD sample, while the output WGs are spanning over the rest of the PIC sample towards the detection setup. 
To reduce the distance between the WG ends and the edges of the Si$_3$N$_4$ sample as much as possible, considering the dimensions of the DLW structures, a subsequent dicing step was performed.
After finishing the individual fabrication of the III-V and Si$_3$N$_4$ samples, both were manually aligned to each other on top of a silicon-carbide (SiC) carrier. A proper alignment was achieved by using etched markers in the Si$_3$N$_4$ layer as reference positions. By backside illumination through the SiC a UV-curable adhesive secured the positioning, a subsequent development, using PGMEA (propylene glycol methyl ether acetate), removed excess adhesive that might have ended up on top of the samples.
\newline
In a last fabrication step the PWBs were created, connecting the III-V platform with the Si$_3$N$_4$-based PIC. Additionally, the lensed side couplers at the outputs of the PIC were fabricated. The latter, shown in Figure \ref{fig:chip_figure}b), are suitable to enhance the coupling efficiencies in the given off-chip detection configuration. They were the first 3D components fabricated due to their higher mechanical stability compared to the PWBs. For this a Nanoscribe GT system was used in combination with the high refractive index polymer IP-n162 ($n>1.59$ at \SI{1550}{\nano\meter}). By alignment through image detection onto etched markers, schematically shown in Figure \ref{fig:Excitation_Sketch}, the side couplers were positioned on the WG ends. They increase the WG mode by using a taper, which increases linearly over \SI{230}{\micro \meter} from a square cross section with a side length of \SI{1.5}{\micro\meter} to \SI{14}{\micro\meter}. The enlarged mode is met by a spherical lens with a radius of \SI{28}{\micro \meter}, to compensate the divergence of the outcoupled beam. After a subsequent development, the sample-system was again mounted in a Nanoscribe GT system for the fabrication of the PWBs. They are created from the polymer IP-Dip and aligned towards the integrated Si$_3$N$_4$ photonics by using the same marker detection as mentioned before. An overview of the PWB structure can be seen in Figure \ref{fig:chip_figure}c). To couple the light from the PWBs into the WGs, a mode converter, as already presented similarly \cite{Gehring:2019_2}, is used. It consists of an inverse tapered Si$_3$N$_4$ section, that is embedded in a rectangular polymer WG of constant width, and increasing height. To avoid printing artefacts at the rough edge of the Si$_3$N$_4$ sample due to the dicing process, the PWBs were lifted before they reach the edge of the sample. After a bend towards the QD sample, the PWBs are tapered from their original cross section of 2x\SI{2}{\micro \meter^2} to 3x\SI{3}{\micro \meter^2} to enhance the mechanical stability. 
The development of the PWBs was performed with PGMEA and isopropanol, with an additional step of Novec\textsuperscript{TM} (Sigma-Aldrich SHH0002) to avoid the necessity of blow drying the sample.

\section{Experimental results}
\begin{figure*}[hbt!]
\includegraphics{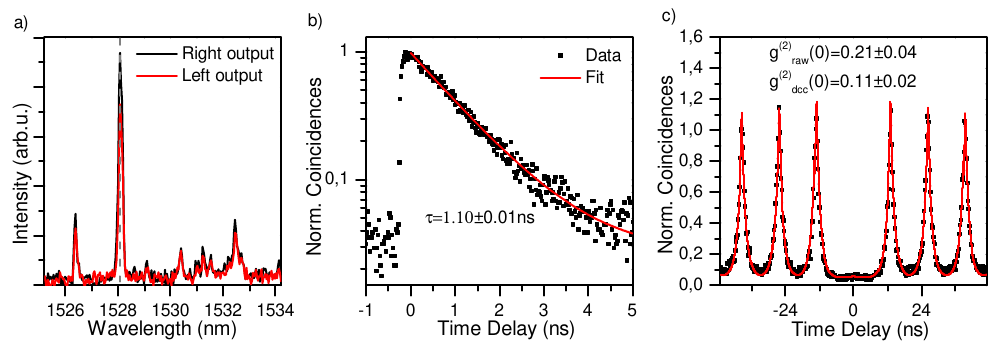}
\caption{\label{fig:wide}a) Measured spectra of both beamsplitter output ports of the investigated QD. The dashed grey line indicates the investigated transition. b) Life time measurement of the investigated transition at \SI{1528.1}{\nano\meter} under weak pulsed AB excitation. c) Second-order correlation measurement under pulsed AB excitation at saturation power with a binwidth of \SI{250}{\pico \second}. The $g^{(2)}_\mathrm{raw}(0)$ was determined by comparing the integrated coincidences in a \SI{13.14}{\nano\second} window around the zero-delay peak and the average coincidences in the same window around a peak in the Poissonian level. $g^{(2)}_\mathrm{dcc}(0)$ is the value corrected for the additional coincidences due to the SNSPD dark counts.}
\end{figure*}
The fully assembled sample-system was mounted in a helium-flow cryostat allowing the excitation from the top and the simultaneous collection of the horizontal emission from the side couplers, as schematically shown in Figure \ref{fig:Excitation_Sketch}. For the collection of the emitted light, both output ports of the MMI were used with the help of an objective with a NA of \num{0.6}. Both emission paths were spatially separated by a 4f setup and afterwards coupled into single-mode fibers. 
Several structures suitable for further experiments with a splitting ratio (SR) close to 50:50 (R:T) were found.
For multiple of these structures, the emission spectrum showed different transition lines compared to the bare pre-selected microlenses. 
This behavior is attributed to non ideal collection condition due to the non-deterministic mircolens positioning,
together with the spatially wide collection window of the \SI{3}{\micro\meter} broad PWB interface. This allows nearby QDs emission to couple and may change the collected spectra. This issue can be reduced in the future when deterministic fabrication is employed and the centered QD dominates the present emission.  
Furthermore, a slight blue shift ($\approx$\SI{0.84}{\milli \eV}) was also observed on the pre-selected lines which is attributed to the induced strain by the DLW structures during the cooling down as well as to the changed excitation conditions due to the presence of the PWB structures \cite{Sartison2017}.
In the following, one of the investigated structures showing stable emission over several cooling cycles is investigated in more detail. This structure shows a SR of $(57.5:42.5)\pm$\SI{0.6}{\percent}, both detected output spectra are shown in Figure \ref{fig:wide}a). The QD displays a main transition line, attributed to a trion, at \SI{1528.1}{\nano\meter}. 
This transition was excited using weak AB excitation at \SI{800}{\nano\meter} with a \SI{76}{\mega\hertz} excitation rate. The emission was then spectrally filtered using a tunable fiber coupled bandpass filter with a full width at half maximum of \SI{0.5}{\nano\meter} and time-correlated single photon counting was performed. The measured decay time $\tau$ in Figure \ref{fig:wide}b) corresponds to \SI{1.10(0.01)}{\nano\second}, which is in the expected range for QDs grown on a MMB \cite{Sittig2022,Paul2017}. To demonstrate that single-photons were successfully coupled from the III-V sample to the Si$_3$N$_4$-based PIC, both MMI outputs were filtered and detected simultaneously. The MMI is used as a beamsplitter in a Hanbury Brown and Twiss interferometer, allowing to measure the second-order correlation function by detecting the single-photons with superconducting nanowire single-photon detectors (SNSPDs). Under pulsed AB excitation in saturation, a nearly vanishing center peak in Figure \ref{fig:wide}c) proves single-photon operation. The data analysis shows an integrated value of $g^{(2)}_\mathrm{raw}(0)=\SI{0.21(0.04)}{}$ for a time window of \SI{13.14}{\nano\second}. The uncorrelated background caused by the SNSPD dark counts of \SI{300}{\hertz} can be corrected, resulting in $g^{(2)}_\mathrm{dcc}(0)=\SI{0.11(0.02)}{}$. The non-vanishing contribution in the $g^{(2)}(0)$ is attributed to residual uncorrelated background emission due to the utilized AB excitation. In the future, this could be improved by using quasi-resonant or resonant excitation schemes, like p-shell excitation or the SUPER scheme \cite{Bracht2021}. From the blinking behavior \cite{Santori2001}, an optical on-time of $\beta_\mathrm{on}=\SI{80(0.01)}{\percent}$ was estimated. 
The measurements were performed with a dark count subtracted count rate of $\approx 4600$ counts per second and per output, corresponding to an overall efficiency of $\eta_\mathrm{Total}=\SI{0.012}{\percent}$. This is comparable to another PWB related work with QDs as single-photon source \cite{Gregorio2024}.
The overall efficiency is here strongly limited by the microlens, which provides a more Gaussian-like emission profile according to simulations, but showed experimentally no increase in the extraction efficiency compared to planar samples with a weak $\lambda$-cavity between bottom DBR and the semiconductor-vacuum interface.

\section{Interface performance}
\begin{figure}[hbt!]
\includegraphics[width=0.45\textwidth]{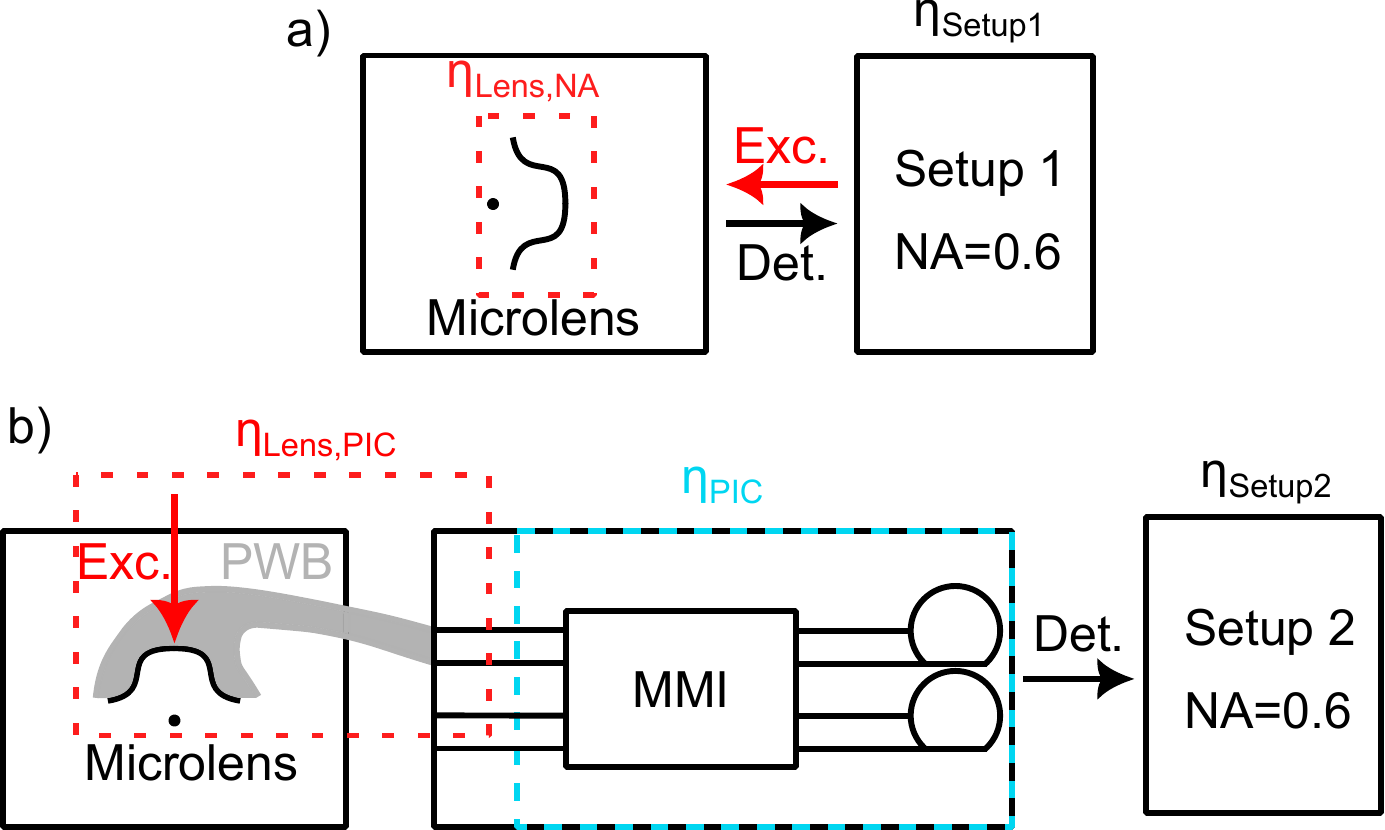}
\caption{\label{fig:eff_skizze_v2} Schematic illustration of the different measurement configurations: a) Setup 1 used for the pre-selection of the microlenses allowing a confocal excitation and detection. b) Setup 2 used for the investigation of the fully assembled hybrid chip enabling orthogonal excitation and detection at the same time.}
\end{figure}
The successful transfer of single-photons into the PIC has been demonstrated and the efficiency of the hybrid approach was further analyzed. Particularly the PWB efficiency plays a fundamental role for the presented hybrid approach.
A direct efficiency measurement of the PWB transmission was not feasible since the microlenses on the III-V sample also have to be considered in the transmission definition. Therefore, the ratio $\Gamma_\mathrm{trans}$ between the lens to PIC transmission efficiency $\eta_\mathrm{Lens,PIC}$ and the detection efficiency $\eta_\mathrm{Lens,NA}$ of the bare microlenses using an NA of \num{0.6} is quantified by isolating both values via
\begin{equation} \label{eq:3D_efficiency_v2}
    \Gamma_\mathrm{trans} = \frac{\eta_\mathrm{Lens,PIC}}{\eta_\mathrm{Lens,NA}} = \frac{\overline{I_\mathrm{PIC}}/(\eta_\mathrm{PIC}\eta_\mathrm{Setup2})}{\overline{I_\mathrm{Lens}}/\eta_\mathrm{Setup1}}
    .
\end{equation}
For the comparison, the average intensity of 9 working PWB structures $\overline{I_\mathrm{PIC}}$ and of the 52 pre-selcted bare microlenses $\overline{I_\mathrm{Lens}}$ are determined by summing up the measured spectra. This has the advantage of taking several working structures into account, providing a more precise estimation of the general performance of the presented approach. Different measurement configurations were used due to the varying excitation and detection geometries and therefore have to be taken into account, as schematically shown in Figure \ref{fig:eff_skizze_v2}. The efficiency of the used setups were measured, corresponding to $\eta_\mathrm{Setup1}= \SI{41.8(0.6)}{\percent}$ for the pre-selection and $\eta_\mathrm{Setup2}= \SI{37.8(0.3)}{\percent}$ for the hybrid approach. For the latter also the PIC efficiency $\eta_\mathrm{PIC}$ has to be considered. The losses directly related to the Si$_3$N$_4$-based PIC elements can be measured individually and are only minimal, with \SI{98.0(0.1)}{\percent} transmission efficiency for the full WG length and additional \SI{94.8(1.6)}{\percent} for the MMIs. 
More dominant are the side coupler losses, caused by the mode-converter section, internal reflections at the polymer to air interface and a slight mode mismatch reducing the achievable fiber coupling, resulting in an efficiency of \SI{66.4 \pm 2.4}{\percent}. Overall, the PIC shows a total transmission of 
$\eta_\mathrm{PIC}= \SI{61.7(3.4)}{\percent}$. 
Inserting all values in Equation \eqref{eq:3D_efficiency_v2} to isolate $\eta_\mathrm{Lens,NA}$ and $\eta_\mathrm{Lens,PIC}$ results in a transfer ratio of $\Gamma_\mathrm{trans}=\SI{28.6(8.8)}{\percent}$ for the microlens to PIC interface compared to the bare microlens in combination with an objective with an NA of \num{0.6}. It can be seen that the amount of transferred photons is clearly smaller than the one measured for the bare microlens. We note that a higher collection effiency at the PWB to microlens interface is expected due to the spatially wide collection range compared to the limited NA of the employed objective. The nevertheless reduced transmission efficiency can be attributed to the bending and DLW imperfection of the PWBs, introducing losses on the guided signal. The dimensions of the PWBs, in terms of length and diameter, also introduce additional losses. The necessary printing distance of $\approx \SI{200}{\micro\meter}$ and the different expansion coefficients of the material platforms increases the mechanical stress during the cooling process. 
The diameter of the PWBs (2x$\SI{2}{\micro \meter^2}$) is large enough to support multiple modes and was chosen for mechanical stability than for optical reasons. While this in principle allows a more efficient collection of the light emitted by the QDs, it reduces the overlap to the single-mode Si$_3$N$_4$ WGs. Especially the discrimination of TM components should suppress the efficiency significantly as it is to be expected that the QDs emit more or less uniformly in TE and TM parts in AB excitation.
%, which then indicates that the TM fundamental mode should carry as much intensity as the one that is TE polarized. 
Thus, an increased efficiency is achievable in future samples using polarization sensitive WGs. 
Also artifacts, stemming from standing waves which occur at the III-V surface due to the high reflectivity, as reported already \cite{Perez2023}, may lead to additional losses.
\section{Summary and outlook}
In conclusion, single-photons from In(Ga)As QDs  were successfully funneled into a Si$_3$N$_4$-based PIC by employing two-photon polymerization to create PWBs connecting both platforms. Truncated Gaussian-shaped microlenses were used to achieve a more Gaussian-like emission profile to increase the coupling efficiency into the PWB. For the PWBs an average transfer ratio of $\Gamma_\mathrm{trans}=\SI{28.6(8.8)}{\percent}$ directly into the SI$_3$N$_4$-based PIC was determined in comparison to the bare microlenses employing an NA of 0.6.
The photons were guided into a beamsplitter forming a Hanbury-Brown and Twiss interferometer and a dark count corrected $g^{(2)}_\mathrm{corr}(0)$ of $\SI{0.11(0.02)}{}$ was measured. This proves the effective realization of a hybrid PIC with a III-V-based single-photon source and Si$_3$N$_4$-based photonics.
For further experiments, other excitation schemes like p-shell excitation, phonon assisted excitation or the SUPER scheme \cite{Bracht2021}, which has been already demonstrated for the investigated kind of QDs \cite{Joos2024}, can be used to approach a higher single-photon purity. 
The overall efficiency from the QDs to the Si$_3$N$_4$ circuitry could be increased by further optimizing the PWBs in terms of surface roughness as well as the coupling to the integrated WGs. The overall brightness can be improved by implementing other surface emitting architectures like circular Bragg gratings for an enhanced extraction efficiency and a Purcell enhancement of the QD emission \cite{Nawrath2023}.
Additionally, the device efficiency can be enhanced by reducing the necessary bending, which can be achieved by replacing the microlens assisted QDs with well established III-V based ridge WGs, as it has been demonstrated with In(Ga)As QDs operating at \SI{900}{\nano\meter} \cite{KDJöns2015} and with droplet etched GaAs QDs emitting at around \SI{780}{\nano\meter} \cite{Hornung2024}.
The proof-of-principle measurements in this work shows a method to combine the III-V platform with the highly appealing Si$_3$N$_4$ platform via the DLW technology. This can be used in the future to design more complex systems with an increased amount of optical elements being integrated on chip, like filters \cite{FBP21}, modulators and on-chip detectors \cite{Beutel:22}.
\section{Acknowledgments}
We acknowledge the financial support of the Deutsche Forschungsgemeinschaft (DFG) via the Project 469373712, the Project GRK2642 and the CRC1459. We also thank the German Federal Ministry of Education and Research (BMBF) for the support via Project QR.X (16KISQ013). Additionally, we acknowledge the support by the European Union’s Horizon 2020 research and innovation program (grant no. 101017237, PHOENICS project). We thank Jonas Schütte for great help with the photograph in Figure \ref{fig:chip_figure}. We acknowledge the Münster Nanofabrication Facility (MNF) for providing the tools used in the polymer and Si$_3$N$_4$ fabrication. We warmly acknowledge the strong support of Michal Vyvle\v{c}ka in scientific discussions.
\section{Author contributions}
U.P. perfomed the optical and quantum optical measurements with the support of E.H., R.J. and F.H.. D.W. designed and prepared the Si$_3$N$_4$ sample and performed the $3$D printing with support from E.J.. D.W. and H.H. developed the side couplers. L.E. realized the microlenses, while P.V. grew the sample with the support of M.J.. U.P. and D.W. wrote the manuscript with support from S.L.P.. U.P. and D.W. analysed the data. S.L.P., W.H.P. and P.M. designed the experiment and coordinated the project. All authors contributed to the revision of the manuscript and scientific discussions.
\section{Data availability}
All data needed to evaluate the conclusions in the paper are present in the main text. Raw data are available from authors upon reasonable request.
\section{Competing interests statement}
All authors declare no financial or non-financial competing interests.

\subsection{References}
\bibliographystyle{nature}%{unsrtnat-AQT}
\bibliography{bibtex}

\end{document}